# Algorithms for Discrete Denoising Under Channel Uncertainty

George M. Gemelos   Styrmir Sigurjónsson   Tsachy Weissman

**Abstract**

The goal of a denoising algorithm is to reconstruct a signal from its noise-corrupted observations. Perfect reconstruction is seldom possible and performance is measured under a given fidelity criterion. In a recent work, the authors addressed the problem of denoising unknown discrete signals corrupted by a discrete memoryless channel when the channel, rather than being completely known, is only known to lie in some uncertainty set of possible channels. A sequence of denoisers was derived for this case and shown to be asymptotically optimal with respect to a worst-case criterion argued most relevant to this setting. In the present work we address the implementation and complexity of this denoiser for channels parametrized by a scalar, establishing its practicality. We show that for symmetric channels, the problem can be mapped into a convex optimization problem, which can be solved efficiently. We also present empirical results suggesting the potential of these schemes to do well in practice. A key component of our schemes is an estimator of the subset of channels in the uncertainty set that are feasible in the sense of being able to give rise to the noise-corrupted signal statistics for some channel input distribution. We establish the efficiency of this estimator, both algorithmically and experimentally. We also present a modification of the recently developed discrete universal denoiser (DUDE) that assumes a channel based on the said estimator, and show that, in practice, the resulting scheme performs well. For concreteness, we focus on the binary alphabet case and binary symmetric channels, but also discuss the extensions of the algorithms to general finite alphabets and to general channels parameterized by a scalar.

## I. INTRODUCTION

In [10] it was shown that optimum denoising of a finite-alphabet process corrupted by a discrete memoryless channel (DMC) whose associated transition matrix is invertible can be achieved, asymptotically, without knowledge of the source statistics, provided the channel is known. In [4] the problem where, in addition to the lack of knowledge of the source statistics, there is also uncertainty in the channel characteristics was addressed. Motivation for the setting of channel uncertainty is also discussed in [4]. The main focus of this paper is on algorithms for implementing the denoising schemes suggested in [4] for channels that can be parametrized by a scalar. An example of such a channel is a binary symmetric channel (BSC), parametrized by the crossover probability, which uniquely defines the channel. We shall focus for concreteness on the case of binary alphabets, since this is enough to capture the essence of the problem while minimizing cumbersome notation. We then demonstrate how the algorithms extend to the case of non-binary alphabets for channels parametrized by a scalar. We shall also make some observations regarding fundamental differences between the denoisers suggested in [4] and the Discrete Universal DEnoiser (DUDE) of [10]. In particular, we show that the suggested denoiser is not a special case of the DUDE, and that the DUDE will in general be suboptimal under the performance criteria of [4].

Authors are with the department of electrical engineering, Stanford University, Stanford, CA 94305.
The work of the first two authors was supported by MURI Grant DAAD-19-99-1-0215 and NSF Grants CCR-0311633 and CCF-0512140. The work of the third author was supported in part by NSF Grant CCR-0312839.

In the setting of denoising for a known channel [10], there is a one to one correspondence between the channel output distribution and its input distribution. Our present setting is fundamentally different in that, given a noise-corrupted process, there may be many source-channel pairs that can give rise to it. As an example consider the binary symmetric channel. If the output process is a Bernoulli($\alpha$) process, there are uncountably many source-channel combinations which can produce the output distribution, e.g., both a Bernoulli(0) process going through a BSC with crossover probability $\alpha$, and a Bernoulli($\alpha$) process going through a noise-free BSC (with crossover probability 0). This is a basic difference which renders the task of attaining the performance of the optimum non-universal Bayesian scheme impossible, even for a scheme with complete knowledge of the noise-corrupted signal statistics. This point is elaborated on in [4].

It has thus been argued in [4] that, under these circumstances, given any noise-corrupted source, a natural criterion under which the performance of a denoising scheme should be judged is its worst case performance under all source-channel pairs consistent with the given noise-corrupted source distribution. This clearly bounds the attainable worst case performance since, in our universal setting, even the noise-corrupted signal distribution is not a priori known.

A family of denoisers that are universally optimal in the sense of asymptotically minimizing the said worst case performance are presented in [4]. The main contribution of the present work is in establishing the practicality of the denoisers of [4], by developing low complexity algorithms for implementing these denoising rules, in the case where the channel is parametrized by a scalar.

When considering the binary alphabet case we shall assume the setting of a binary signal with an unknown distribution, corrupted by a BSC with a crossover probability $\delta$ known to satisfy $\delta < 1/2$, but otherwise unknown. As will be detailed, our work in the binary alphabet setting conveys the essence of the implementation of the scheme for an arbitrary finite alphabet and an invertible channel matrix that depends on a single parameter, only known to lie in a given uncertainty set.

After introducing some notation in Section II, we turn in Section III to the binary minimax setting, where the goal is to minimize the maximum expected fraction of errors made by the denoiser over all source-channel pairs consistent with the noise-corrupted source distribution, where the channel also lies in the uncertainty set. In this section we present the denoiser suggested in [4], as well as one of its performance guarantees. We then present our algorithm for efficiently implementing the suggested denoiser. With careful analysis the problem can be mapped into three convex optimization problems. Each problem can be solved efficiently, and the denoiser achieving the minimum can be found.

In Section IV we discuss a key component of the algorithm, namely a method for estimating the set of "feasible" channels. For a given noise-corrupted source, we say that a channel is feasible if there exists a clean source such that when passed through the channel can give rise to the given noise-corrupted source. In particular, we present a low complexity algorithm to efficiently estimate the set of feasible channels, and compare its performance to a scheme suggested in [10] on binary symmetric channels. In Section V we explicitly describe the binary minimax denoiser via its pseudo-code. Furthermore, we analyze the complexity of the algorithm. In particular, we show that the suggested minimax denoiser has the same order of complexity as the DUDE of [10], namely linear in the size of the data.

We present experimental results in Section VI. We compare the performance of our algorithm, to estimate the set of feasible channels, to that of an algorithm suggested in [10]. We report the results of several simulations, comparing the performance of our denoiser with various versions of the DUDE of [10]. We show that the minimax denoiser performs well, in many cases outperforming the DUDE using various estimates of the channel parameter. However, when the estimate of the channel parameter is accurate, the DUDE using that estimate will perform close to optimally. In particular, it may do better than the minimax denoiser whose performance is inherently more conservative. The simulations include both one- and two-dimensionally indexed data.

In Section VII we demonstrate ways to extend the algorithms of Sections III and IV, which were developed for the binary case, to the more general case of non-binary alphabets and channels parameterized by a scalar, with naturally structured channel uncertainty sets. We briefly discuss the general case of multi-alphabet implementation for general channels that can be appropriately parametrized by a scalar, and describe a method for solving the optimization problem that arises.

In Section VIII we show that the min-max in our problem setting is not equivalent to the max-min, implying the minimax denoiser is not, in general, a special form of the DUDE of [10]. We conclude in Section IX with a summary of our results.

## II. DEFINITIONS AND NOTATION

We assume that the components of the noise-free signal, the observation and the reconstruction signal take their values in the same finite alphabet $\mathcal{A}$. Let $\mathcal{C}(\mathcal{A})$ denote the set of all invertible Discrete Memoryless Channels (DMCs)[1] with input and output alphabet $\mathcal{A}$. We identify an element $\Pi \in \mathcal{C}(\mathcal{A})$ with a $|\mathcal{A}| \times |\mathcal{A}|$ stochastic matrix, $\Pi(a, b)$ denoting the probability of a symbol $b$ at the channel output when the channel input is $a$. Let $\mathcal{S}(\mathcal{A})$ denote the simplex of probability distributions on $\mathcal{A}$.[2] Denote by $\Lambda : \mathcal{A} \times \mathcal{A} \to \mathbb{R}$ a given loss function where $\Lambda(x, \hat{x})$ is the loss incurred when estimating the symbol $x$ with the symbol $\hat{x}$. A randomized $n$-block denoiser $\hat{X}^n$ is a mapping $\hat{X}^n : \mathcal{A}^n \to (\mathcal{S}(\mathcal{A}))^n$, i.e. upon seeing the noise-corrupted signal $z^n \in \mathcal{A}^n$, the $i$-th reconstruction symbol is $a \in \mathcal{A}$ with probability $\hat{X}^n_{[i]}(z^n)[a]$, where $\hat{X}^n_{[i]}(z^n) \in \mathcal{S}(\mathcal{A})$ denotes the $i$-th component of $\hat{X}^n(z^n)$ and $\hat{X}^n_{[i]}(z^n)[a]$ denotes the probability that it assigns to $a \in \mathcal{A}$. The construction of randomized denoisers is motivated by the minimax performance measure which is introduced in Section III.

Uppercase letters will denote random quantities while lower case letters denote deterministic values. Bold notation will indicate doubly infinite sequences e.g. $\mathbf{X} = (\ldots, X_{-1}, X_0, X_1, \ldots)$. Also, $X_i^j$ denotes the sequence $(X_i, \ldots, X_j)$, omitting the subscript when $i = 1$. For $x^n \in \mathcal{A}^n$ and $z^n \in \mathcal{A}^n$ we denote

$$L_{\hat{X}^n}(x^n, z^n) = \frac{1}{n} \sum_{i=1}^n \sum_{a \in \mathcal{A}} \Lambda(x_i, a) \hat{X}^n_{[i]}(z^n)[a]. \quad (1)$$

In words $L_{\hat{X}^n}(x^n, z^n)$ is the expected normalized cumulative loss of the denoiser $\hat{X}^n$ on the individual sequence pair $(x^n, z^n)$, where the expectation is with respect to the randomization of the denoiser.

---
[1]A channel is invertible if the inverse of its transition matrix exists.
[2]Similarly, we will use $\mathcal{S}^k(\mathcal{A})$ to denote the simplex on $k$-tuples on the alphabet $\mathcal{A}$. Also, $\mathcal{S}^\infty(\mathcal{A})$ will denote the set of all distribution on doubly-infinite sequences that take value in $\mathcal{A}$.

With the exception of Section VII, we consider processes with binary alphabets i.e. $\mathcal{A} = \{0, 1\}$, corrupted by a BSC and take $\Lambda(\cdot, \cdot)$ to be the Hamming loss function. In this case a $n$-block denoiser $\hat{X}^n$ is a mapping $\hat{X}^n : \{0,1\}^n \to [0,1]^n$, $\hat{X}^n_{[i]}(z^n)[1] \in [0,1]$ denoting the probability that the $i$-th reconstruction symbol is a $1$ when the denoiser observes $z^n$. For $x^n \in \{0,1\}^n$ and $z^n \in \{0,1\}^n$ (1) becomes

$$L_{\hat{X}^n}(x^n, z^n) = \frac{1}{n} \sum_{i=1}^{n} \left| x_i - \hat{X}^n_{[i]}(z^n)[1] \right|. \tag{2}$$

Let $\mathcal{F}_k \triangleq \{f : \mathcal{A}^{2k+1} \mapsto \mathcal{S}(\mathcal{A})\}$ be the set of all $k$-order sliding window denoisers. A sliding window denoiser of order $k$ works as follows: When denoising a particular symbol it considers the $k$ symbols preceding it and the $k$ symbols succeeding it. These $k$ symbols before and after the current symbol, form a double-sided context of the current symbol. In particular, if we denote the current symbol by $z_0$, the two-sided context is $z_{-k}^{-1}$ and $z_1^k$.

For a given denoiser $f \in \mathcal{F}_k$, we use $f(z_{-k}^k)[a]$ to denote the probability of reconstructing $z_0$ with the letter $a$. With this in mind, we define

$$L_f(x^n, z^n) = \frac{1}{n} \sum_{i=1}^{n} \left| x_i - f(z_{i-k}^{i+k})[1] \right|, \tag{3}$$

namely, the expected normalized loss when employing the sliding-window denoiser defined by $\hat{X}^n_{[i]}(z^n) = f(z_{i-k}^{i+k})$.

### III. THE MINIMAX CRITERION

Our setting assumes uncertainty in the channel characteristics. We quantify this uncertainty by assuming that we are *given* a set of channels which includes the true channel that corrupted the clean source of interest. In the binary setting we assume the uncertainty set to consist of BSCs. The set is parameterized by the crossover probability and, for the sake of simplicity, we assume it is of the form $[0, \mathcal{U}]$, $\mathcal{U} < 1/2$.

For a stationary ergodic source $P_{\mathbf{Z}}$ let

$$\mathcal{C}_\infty(P_{\mathbf{Z}}) = \{0 \leq \delta \leq 1/2 : \exists P_{\mathbf{X}} \text{ s.t. } P_{\mathbf{X}} * \delta = P_{\mathbf{Z}}\}, \tag{4}$$

where $P_{\mathbf{X}} * \delta$ denotes the output distribution of a BSC($\delta$) whose input process has distribution $P_{\mathbf{X}}$. Thus when the output process has distribution $P_{\mathbf{Z}}$ the possible source-channel pairs can be described by the following set, $\{(P_{\mathbf{X}}, \delta) : P_{\mathbf{X}} * \delta = P_{\mathbf{Z}}, \delta \in \mathcal{C}_\infty(P_{\mathbf{Z}})\}$. Further we let

$$\Gamma(P_{\mathbf{Z}}) = \max \mathcal{C}_\infty(P_{\mathbf{Z}}). \tag{5}$$

It is easy to see that $\mathcal{C}_\infty(P_{\mathbf{Z}}) = [0, \Gamma(P_{\mathbf{Z}})]$, since if $\delta \in \mathcal{C}_\infty(P_{\mathbf{Z}})$ then $\delta' \in \mathcal{C}_\infty(P_{\mathbf{Z}})$. for any $\delta' < \delta$. Define

$$\Delta(P_{\mathbf{Z}}) = \min\{\mathcal{U}, \Gamma(P_{\mathbf{Z}})\}.$$

Hence, $[0, \Delta(P_{\mathbf{Z}})]$ is the set of all channels in the uncertainty set that can give rise to $P_{\mathbf{Z}}$ when corrupting some noiseless source. Note that $\Delta$ only depends on $P_{\mathbf{Z}}$ through $\Gamma$ and this dependence will often be omitted.

For an $n$-block denoiser $\hat{X}^n$ let

$$\mathcal{L}^{(n)}_{\hat{X}^n}(P_{\mathbf{Z}}, \Delta, \mathbf{Z}) = \sup_{\{(P_{\mathbf{X}}, \delta) : P_{\mathbf{X}} * \delta = P_{\mathbf{Z}}, \delta \in [0, \Delta] \cap \mathbb{Q}\}} E_{P_{\mathbf{X}}, \delta} \left[ L_{\hat{X}^n}(X^n, Z^n) | \mathbf{Z} \right], \tag{6}$$

where $E_{P_{\mathbf{X},\delta}}[\cdot|\mathbf{Z}]$ denotes expectation conditioned on $\mathbf{Z}$ under the joint distribution on the noiseless and noise-corrupted source induced when the noiseless source $\sim P_{\mathbf{X}}$ is corrupted by a BSC($\delta$). Thus $\mathcal{L}^{(n)}_{\hat{X}^n}(P_{\mathbf{Z}}, \Delta, \mathbf{Z})$ is the worst case performance of the denoiser $\hat{X}^n$ over all channels that can give rise to $P_{\mathbf{Z}}$, given the noisy realization $\mathbf{Z}$. In the definition of $\mathcal{L}^{(n)}_{\hat{X}^n}(P_{\mathbf{Z}}, \Delta, \mathbf{Z})$ we take the sup over the set of all channels in $[0, \Delta]$ intersected with the rational numbers, $\mathbb{Q}$. This is to guarantee that the sup is over a countable set and thus that $\mathcal{L}^{(n)}_{\hat{X}^n}(P_{\mathbf{Z}}, \Delta, \mathbf{Z})$ is a well-defined random variable. The rationale of using (6) as the performance measure in the unknown channel setting is discussed in [4].

Define now

$$\mu^{(n)}_k(P_{\mathbf{Z}}, \Delta, \mathbf{Z}) = \min_{f \in \mathcal{F}_k} \mathcal{L}^{(n)}_f(P_{\mathbf{Z}}, \Delta, \mathbf{Z}), \tag{7}$$

Further let

$$\mu_k(P_{\mathbf{Z}}, \Delta, \mathbf{Z}) = \limsup_{n \to \infty} \mu^{(n)}_k(P_{\mathbf{Z}}, \Delta, \mathbf{Z}). \tag{8}$$

Therefore $\mu_k$ is the performance of the best sliding window denoiser of length $2k + 1$. Building on the definition of $\mu_k$ let the "sliding window minimum loss" be defined by

$$\mu(P_{\mathbf{Z}}, \Delta, \mathbf{Z}) = \lim_{k \to \infty} \mu_k(P_{\mathbf{Z}}, \Delta, \mathbf{Z}), \tag{9}$$

where the limit exists since $\mu_k(P_{\mathbf{Z}}, \Delta, \mathbf{Z})$ is clearly non-increasing with $k$. Since $\mu_k$ is the performance of the best sliding window denoiser of length $2k + 1$, $\mu$ is the performance of the best finite length sliding window denoiser.

## A. Construction of a Universal Scheme

For $\alpha \in [0, 1]$, $\delta < 1/2$, $d_0 \in [0, 1]$ and $d_1 \in [0, 1]$ let

$$F(\alpha, \delta, d_0, d_1) = \frac{1}{1 - 2\delta} \left[(1 - \delta - \alpha)(1 - \delta)d_0 + (1 - \delta - \alpha)\delta d_1 + (\alpha - \delta)\delta(1 - d_0) + (\alpha - \delta)(1 - \delta)(1 - d_1)\right]. \tag{10}$$

In the single observation problem, for $\alpha \in [\delta, 1 - \delta]$, $F(\alpha, \delta, d_0, d_1)$ is the expected loss of a denoising scheme which says 1 with probability $d_0$ upon observing a zero and says 1 with probability $d_1$ upon observing a one, when observing a Bernoulli($(\alpha - \delta)/(1 - 2\delta)$) corrupted by a BSC($\delta$). Note that in such a case, the channel output is a Bernoulli($\alpha$). For a probability distribution on binary $(2k + 1)$-tuples, $P_{Z^k_{-k}}$, and $f : \{0, 1\}^{2k+1} \to [0, 1]$ we define the functional $G_k$ by

$$G_k\left(P_{Z^k_{-k}}, \delta, f\right) = \sum_{z^{-1}_{-k}, z^k_1 \in \{0,1\}^k} F\left(P_{Z_0|z^{-1}_{-k},z^k_1}(Z_0 = 1), \delta, f([z^{-1}_{-k}, 0, z^k_1]), f([z^{-1}_{-k}, 1, z^k_1])\right) P_{Z^k_{-k}}(z^{-1}_{-k}, z^k_1), \tag{11}$$

where $P_{Z_0|z^{-1}_{-k},z^k_1}$ denotes $\Pr(Z_0 = 1|Z^{-1}_{-k} = z^{-1}_{-k}, Z^k_1 = z^k_1)$ under the source $P_{Z^k_{-k}}$ and $[z^{-1}_{-k}, z_0, z^k_1]$ denotes the $(2k + 1)$-tuple. We further define $J_k$ by

$$J_k\left(P_{Z^k_{-k}}, \Delta, f\right) = \max_{\delta \in [0, \Delta]} G_k\left(P_{Z^k_{-k}}, \delta, f\right), \tag{12}$$

and

$$f_{\mathrm{MM}_k}\left(P_{Z^k_{-k}}, \Delta\right) = \arg\min_{f \in \mathcal{F}_k} J_k\left(P_{Z^k_{-k}}, \Delta, f\right), \tag{13}$$

selecting an arbitrary achiever when it is not unique.

In our setting $P_\mathbf{Z}$ is not known and thus neither is $\Gamma(P_\mathbf{Z})$ (recall (5) for its definition), which is needed to find $\Delta$, so we need to estimate it. If we knew the distribution of the $l$-tuple $P_{Z^l}$ (induced by $P_\mathbf{Z}$) we could evaluate the following upper bound on $\Gamma(P_\mathbf{Z})$:

$$\Gamma_l\left(P_{Z^l}\right) = \max\left\{0 \leq \delta \leq 1/2 : \exists P_{X^l} \text{ s.t. } P_{X^l} * \delta = P_{Z^l}\right\}. \tag{14}$$

Instead, we shall use the empirical distribution of a $l$-tuple induced by the observation of the noise-corrupted sequence. An efficient algorithm for calculating $\Gamma_l(\cdot)$ is presented in section IV. This leads us to define

$$\Delta_l(P_{Z^l}) = \min\left\{\mathcal{U}, \Gamma_l(P_{Z^l})\right\}. \tag{15}$$

Now let $\hat{Q}^{2k+1}[z^n]$ denote the $(2k+1)$-th order empirical distribution induced by $z^n$ and let $\hat{X}^{n,k,l}$ denote the $n$-block denoiser defined by

$$\hat{X}^{n,k,l}_{[i]}(z^n) = f_{\mathbb{MM}_k}\left(\hat{Q}^{2k+1}[z^n], \hat{\Delta}_l(z^n)\right)\left[z_{i-k}^{i+k}\right] \quad k+1 \leq i \leq n-k, \tag{16}$$

where, with slight abuse of notation we write $\hat{\Delta}_l(z^n)$ as shorthand for $\Delta_l\left(\hat{Q}^l[z^n]\right)$. $\hat{\Delta}_l(z^n)$ is our estimate of the set of feasible channels, i.e., the set of channel crossover probabilities belonging to the uncertainty set that can give rise to the channel output distribution. The denoiser $\hat{X}^{n,k,l}_{[i]}(z^n)$ can be arbitrarily defined for $i$ outside the range $k+1 \leq i \leq n-k$.

## B. Performance Guarantee

We shall now cite one of the results of [4] that provides a sound performance guarantee and therefore justification for the use of the schemes considered in this work. For a sequence $\{\psi_i\}$ of non-negative reals let $\tilde{\mathcal{S}}_{\{\psi_i\}}(\mathcal{A})$ denote the set of stationary distributions whose $i$-th $\psi$-mixing coefficient is upper bounded by $\psi_i$ (cf., e.g., [2], [4] for the definition of $\psi$-mixing coefficients). Most sources arising in practice, such as Markov sources with no restricted sequences and hidden Markov processes with no restricted state sequences can be shown to have exponentially diminishing $\psi$-mixing coefficients [2].

*Theorem 1 (Theorem 2 of [4]):* Let $\{\psi_i\}$ be a sequence of non-negative reals with $\psi_i \to 0$. There exists an unbounded sequences $\{l_n\}$ and $\{k_n\}$ such that if $\hat{X}^n_{univ} = \hat{X}^{n,k_n,l_n}$, where $\hat{X}^{n,k,l}$ was defined in (16), then for any $P_\mathbf{Z} \in \tilde{\mathcal{S}}_{\{\psi_i\}}(\mathcal{A})$, as $n \to \infty$, the performance of $\hat{X}^n_{univ}$ converges to that of the best finite length sliding window denoiser, $\mu(P_\mathbf{Z}, \Delta, \mathbf{Z})$.

## C. Efficient Computation of MiniMax Denoiser

As is evident from (16), the "engine" at the heart of our denoisers is the calculation of $f_{\mathbb{MM}_k}[P_{Z^k_{-k}}, \Delta]$ (defined in (13)), which we now consider. By observing that (10) can be simplified into a function that is quadratic in the maximizing argument, $\delta$, and whose coefficients depend on the minimizing argument, $f$, (11) can be written as,

$$G_k\left(P_{Z^k_{-k}}, \delta, f\right) = \frac{A(P_{Z^k_{-k}}, f)\delta^2 - (1 + A(P_{Z^k_{-k}}, f))\delta + B(P_{Z^k_{-k}}, f)}{1 - 2\delta}, \tag{17}$$

where,

$$A(P_{Z_{-k}^k}, f) = \sum_{c_k} \left(2f(c_k, 0) - 2f(c_k, 1)\right) P_{Z_{-k}^k}(c_k) \tag{18a}$$

$$B(P_{Z_{-k}^k}, f) = \sum_{c_k} \left(\eta_{c_k} + f(c_k, 0) - \eta_{c_k} f(c_k, 0) - \eta_{c_k} f(c_k, 1)\right) P_{Z_{-k}^k}(c_k), \tag{18b}$$

and $\eta_{c_k} = P_{Z_{-k}^k}(Z_0 = 1 | c_k)$. From now on we will suppress $A$ and $B$'s dependence on $P_{Z_{-k}^k}$ and $f$. It is easily shown that (17) is either convex or concave for fixed $A$ and $B$. For the cases it is convex we need to solve,

$$\frac{\partial G_k\left(P_{Z_{-k}^k}, \delta, f\right)}{\partial \delta} = 0. \tag{19}$$

The solutions to (19) are $\delta' = (A + \sqrt{-A^2 - 2A + 4AB})/(2A)$ and $\delta'' = (A - \sqrt{-A^2 - 2A + 4AB})/(2A)$. Therefore to find the maximum in (12), for both the convex and concave case, we have to consider no more than four values for $\delta$, the endpoints of the interval $[0, \Delta]$, $\delta = \delta'$, and $\delta = \delta''$. Also note that $\delta'$ and $\delta''$ only have to be considered if they are feasible, i.e., if they lie in $[0, \Delta]$. Thus (12) is equivalent to

$$\max\left\{B, \frac{A\Delta^2 - A\Delta - \Delta + B}{1 - 2\Delta}, \frac{A\delta'^2 - A\delta' - \delta' + B}{1 - 2\delta'} \mathbb{1}_{\delta' \in (0,\Delta)}, \frac{A\delta''^2 - A\delta'' - \delta'' + B}{1 - 2\delta''} \mathbb{1}_{\delta'' \in (0,\Delta)}\right\}, \tag{20}$$

where $\mathbb{1}$ denotes the indicator function.

To minimize (20) we begin by observing that $A$ and $B$ are linear functions in the minimizing argument, $f$. Since the range of $f$ is convex the observation implies that the range of $(A, B)$ is also convex. Consider now the following modification of (20)

$$\max\left\{B, \frac{A\Delta^2 - A\Delta - \Delta + B}{1 - 2\Delta}, \frac{A\delta'^2 - A\delta' - \delta' + B}{1 - 2\delta'} \mathbb{1}_{\delta' \in (0,\Delta)}\right\}. \tag{21}$$

Define the sets $S(\Delta) = \{(A, B) : \delta'(A, B) \in (0, \Delta)\}$ and $V(\Delta) = \{(A, B) : A \leq (2B - 1)/(1 - \Delta)\}$. The set $S(\Delta)$ is the set of $(A, B)$ corresponding to feasible values of $\delta'$ and the set $V(\Delta)$ is the region where

$$B \geq \frac{A\Delta^2 - A\Delta - \Delta + B}{1 - 2\Delta}.$$

We can then construct the sets $S_1(\Delta) = S(\Delta)^c \cap V(\Delta)$ and $S_2(\Delta) = S(\Delta)^c \cap V^c(\Delta)$. Note that $\{S(\Delta), S_1(\Delta), S_2(\Delta)\}$ is a partition of the range of $(A, B)$ for all $\Delta \in (0, 1/2)$. Furthermore, the constraints

$$\delta' = \frac{A + \sqrt{-A^2 - 2A + 4AB}}{2A} \geq 0$$

$$\delta' = \frac{A + \sqrt{-A^2 - 2A + 4AB}}{2A} \leq \Delta,$$

which define the set $S(\Delta)$, are equivalent to the linear inequalities

$$A \leq 0$$

$$A \leq 2B - 1$$

$$A \geq \frac{2B - 1}{2\Delta^2 - 2\Delta + 1}.$$

Therefore $S(\Delta)$ is an intersection of half-spaces and hence convex. Similarly we can write the set $S_1(\Delta)$ as all the $(A, B)$ such that

$$A \leq \frac{2B-1}{2\Delta^2 - 2\Delta + 1}$$
$$A \leq \frac{2B-1}{1-\Delta}$$

and $S_2(\Delta)$ as all the $(A, B)$ such that

$$A \geq 2B - 1$$
$$A \geq \frac{2B-1}{1-\Delta}.$$

Therefore $S_1(\Delta)$ and $S_2(\Delta)$ are also the intersection of half-spaces and hence convex.

From the construction of $S(\Delta)$, $S_1(\Delta)$, and $S_2(\Delta)$ and the fact that (21) is non-negative we can rewrite (21) as

$$B \mathbb{1}_{S_1(\Delta)} + \frac{A\Delta^2 - A\Delta - \Delta + B}{1 - 2\Delta} \mathbb{1}_{S_2(\Delta)}$$
$$+ \max\left\{ B \mathbb{1}_{S(\Delta)}, \frac{A\Delta^2 - A\Delta - \Delta + B}{1 - 2\Delta} \mathbb{1}_{S(\Delta)}, \frac{A\delta'^2 - A\delta' - \delta' + B}{1 - 2\delta'} \mathbb{1}_{S(\Delta)} \right\}. \tag{22}$$

To further simplify (22), let us examine the max term. The regions where any two of the terms in the max intersect is defined by the following linear equations:

$$A = \frac{2B-1}{1-\Delta} \tag{23a}$$

$$A = 2B - 1 \tag{23b}$$

$$A = \frac{2B-1}{2\Delta^2 - 2\Delta + 1}. \tag{23c}$$

Since all three terms in the maximization are continuous on $S(\Delta)$, the maximizing term can only be exchanged on one of the lines defined in (23). Therefore to find the maximizing term in $S(\Delta)$ we only need to test one point in each region of the partition of $S(\Delta)$ defined by (23). Since equations (23b) and (23c) are on the boundary of the set $S(\Delta)$, we only need to consider the line defined by (23a). It is easy to evaluate the expressions for two points and see that the maximizing term on $S(\Delta)$ is

$$\frac{A\delta'^2 - A\delta' - \delta' + B}{1 - 2\delta'}.$$

We can therefore rewrite (21) as

$$B \mathbb{1}_{S_1(\Delta)} + \frac{A\Delta^2 - A\Delta - \Delta + B}{1 - 2\Delta} \mathbb{1}_{S_2(\Delta)} + \frac{A\delta'^2 - A\delta' - \delta' + B}{1 - 2\delta'} \mathbb{1}_{S(\Delta)}. \tag{24}$$

This leads to the following expression for (20):

$$\max\left\{ B \mathbb{1}_{S_1(\Delta)} + \frac{A\Delta^2 - A\Delta - \Delta + B}{1 - 2\Delta} \mathbb{1}_{S_2(\Delta)} + \frac{A\delta'^2 - A\delta' - \delta' + B}{1 - 2\delta'} \mathbb{1}_{S(\Delta)}, \frac{A\delta''^2 - A\delta'' - \delta'' + B}{1 - 2\delta''} \mathbb{1}_{\delta'' \in (0, \Delta)} \right\}. \tag{25}$$

Now define $S''(\Delta) = \{(A, B) : \delta''(A, B) \in (0, \Delta)\}$, the set of $(A, B)$ corresponding to feasible values of $\delta''$, and rewrite (25) as

$$\left(B\mathbb{1}_{S_1(\Delta)} + \frac{A\Delta^2 - A\Delta - \Delta + B}{1 - 2\Delta}\mathbb{1}_{S_2(\Delta)} + \frac{A\delta'^2 - A\delta' - \delta' + B}{1 - 2\delta'}\mathbb{1}_{S(\Delta)}\right)\mathbb{1}_{S''(\Delta)^c} +$$
$$\max\left\{\left(B\mathbb{1}_{S_1(\Delta)} + \frac{A\Delta^2 - A\Delta - \Delta + B}{1 - 2\Delta}\mathbb{1}_{S_2(\Delta)} + \frac{A\delta'^2 - A\delta' - \delta' + B}{1 - 2\delta'}\mathbb{1}_{S(\Delta)}\right)\mathbb{1}_{S''(\Delta)},\right. \quad (26)$$
$$\left.\frac{A\delta''^2 - A\delta'' - \delta'' + B}{1 - 2\delta''}\mathbb{1}_{S''(\Delta)}\right\}.$$

Similarly to what we did for (22), since the two terms in the maximization are continuous on $S''(\Delta)$, we can look at the region where the two terms intersect. Note that the continuity of both terms in $(A, B)$ follows from that of $\delta'$ and $\delta''$. On $S''(\Delta)$ the two terms intersect only on the lines defined by (23b) and (23c). Similarly in the case of $S(\Delta)$, the lines (23b) and (23c) lie on the boundary of $S''(\Delta)$. Therefore, continuity implies that the maximizing term is not exchanged in $S''(\Delta)$. By picking any point in $S''(\Delta)$ we find that

$$B\mathbb{1}_{S_1(\Delta)} + \frac{A\Delta^2 - A\Delta - \Delta + B}{1 - 2\Delta}\mathbb{1}_{S_2(\Delta)} + \frac{A\delta'^2 - A\delta' - \delta' + B}{1 - 2\delta'}\mathbb{1}_{S(\Delta)}$$

is the maximizing term. Combining this observation and (26) we can rewrite (20) as

$$B\mathbb{1}_{S_1(\Delta)} + \frac{A\Delta^2 - A\Delta - \Delta + B}{1 - 2\Delta}\mathbb{1}_{S_2(\Delta)} + \frac{A\delta'^2 - A\delta' - \delta' + B}{1 - 2\delta'}\mathbb{1}_{S(\Delta)}. \quad (27)$$

Observe that the first two terms in (27) are linear in $(A, B)$ and hence convex.

*Claim 1:*
$$\frac{A\delta'^2 - A\delta' - \delta' + B}{1 - 2\delta'} \quad (28)$$

is convex on $S(\Delta)$ for all $\Delta \in (0, 1/2)$.

**Proof:** The eigenvalues of the Hessian of (28) on the set $S(\Delta)$ are

$$\left\{0, \frac{-A(A + 2 - 4B)\sqrt{4A^2 + 1 + 4B^2 - 4B}}{2A(A + 2 - 4B)^2}\right\}.$$

It is easily shown that the non-trivial eigenvalue has no real roots and is therefore non-zero for all $(A, B) \in S(\Delta)$. Hence, by continuity, to determine whether the Hessian is positive semidefinite on $S(\Delta)$ we only have to evaluate it at a single point. Picking any $(A, B) \in S(\Delta)$, we find that the Hessian is positive semidefinite and, therefore, (28) is convex on $S(\Delta)$. □

Therefore, all three terms in (27) are convex on their respective sets. As shown during their construction, the sets $S(\Delta)$, $S_1(\Delta)$, and $S_2(\Delta)$ are convex sets. Hence the alternative expression (27) tells us that minimizing (20) is equivalent to finding the minimum of three separate convex functions each over a disjoint convex set. This can easily be done by minimizing each term in (20) over its appropriate partition, which can be carried out using efficient convex optimization algorithms. In particular, we have implemented the denoiser using the log barrier method with gradient descent (see [1] for a detailed discussion of the log barrier method). After the three minimizations are carried out it remains only to select the minimum between the three values. Overall, this gives an efficient and concrete way of calculating $f_{\mathbb{MM}_k}[P_{Z_{-k}^k}, \Delta]$. Algorithmic complexity will be discussed in Section V.

## IV. LARGEST CROSSOVER PROBABILITY CONSISTENT WITH OBSERVATIONS

Recall the definitions of $\Gamma(P_{\mathbf{Z}})$ and $\Gamma_l(P_{Z^l})$ (equations (5) and (14)). We shall also use $\Gamma_l(P_{\mathbf{Z}})$ to denote $\Gamma_l$ evaluated for the $l$-th order marginal of $P_{\mathbf{Z}}$. We shall say that $\delta < 1/2$ is $l$-feasible if there exists a distribution on a noise-free tuple $P_{X^l}$ that gives rise to $P_{Z^l}$ when corrupted by a $BSC(\delta)$. Thus, $\Gamma_l$ is the maximum $l$-feasible $\delta$. It is easy to show that for any stationary ergodic $P_{\mathbf{Z}}$, $\Gamma_l(P_{\mathbf{Z}})$ is non-increasing in $l$ and that

$$\Gamma(P_{\mathbf{Z}}) = \lim_{l \to \infty} \Gamma_l(P_{\mathbf{Z}}) = \inf_l \Gamma_l(P_{\mathbf{Z}}). \tag{29}$$

In our setting $P_{\mathbf{Z}}$ is unknown and, consequently, so is $\Gamma(P_{\mathbf{Z}})$. In this section we develop an efficient algorithm to estimate $\Gamma(P_{\mathbf{Z}})$ as a function of the empirical distribution $\hat{Q}^l[z^n]$. Note that this estimate, $\Gamma_l(\hat{Q}^l[z^n])$, or $\hat{\Gamma}_l(z^n)$ for short, is the one implicitly employed also by our denoiser in (16), since the estimate of the channel uncertainty set that it employs $\hat{\Delta}_l$ is taken as the intersection between the a priori channel uncertainty set $[0, \mathcal{U}]$ and the estimated feasible set $[0, \hat{\Gamma}_l]$.

In [4] it is shown that under mild conditions there exists an unbounded sequence $l_n$ such that

$$\lim_{n \to \infty} \hat{\Gamma}_{l_n}(Z^n) = \Gamma(P_{\mathbf{Z}}) \quad P_{\mathbf{Z}} - a.s. \tag{30}$$

In [10, Section 8-C] a method for obtaining an upper bound on $\Gamma_l(P_{\mathbf{Z}})$ is suggested. Let $\{C_j^{(l)}\}_{j=1}^{2^l}$ denote all the $2^l$ binary sequences. The idea is to look at the $\min_{i,j} \varphi_{i,j}^{(2l)}$ where,

$$\varphi_{i,j}^{(2l)} \triangleq \min \left\{ P\left(Z_0 = 1 | Z_{-l}^{-1} = C_i^{(l)}, Z_1^l = C_j^{(l)}\right), P\left(Z_0 = 0 | Z_{-l}^{-1} = C_i^{(l)}, Z_1^l = C_j^{(l)}\right) \right\} \quad \forall i, j, \tag{31}$$

the conditional probability on the right side being the empirical one induced by the data. It is clear that $\min_{i,j} \varphi_{i,j}^{(2l)}$ (for $n$ large enough so that the empirical distribution is close enough to the true one) is an upper bound to $\Gamma_l$. The following is a numerical example illustrating that the method of [10] yields, in general, upper bounds that are not tight, while $\hat{\Gamma}_l$ is guaranteed by (30) to converge to the true value.

*Example 1:* Let $P_{\mathbf{Z}}$ be the first order symmetric Markov Process with transition matrix

$$\begin{pmatrix} 0.695 & 0.305 \\ 0.305 & 0.695 \end{pmatrix}$$

passed through a $BSC(.1)$. For this case clearly $\Gamma(P_{\mathbf{Z}}) \geq 0.1$ and, in fact, the inequality can be shown to be strict [7]. So, in particular, $\Gamma_l(P_{\mathbf{Z}}) > 0.1$ for all $l$. Simulations yield (with high precision and confidence) $\min_{i,j} \varphi_j^{(2)} = 0.2629$, $\min_{i,j} \varphi_j^{(4)} = 0.2440$, $\min_{i,j} \varphi_j^{(6)} = 0.2415$, and $\min_{i,j} \varphi_j^{(8)} = .2398$. On the same simulated data one finds $\hat{\Gamma}_2(z^n) = .1757$, $\hat{\Gamma}_4(z^n) = .1451$, $\hat{\Gamma}_6(z^n) = .1266$, and $\hat{\Gamma}_8(z^n) = .1107$,

### A. Efficient Computation of $\Gamma_l(z^n)$

Our estimate of $\Gamma_l$ simply evaluates $\Gamma_l(\cdot)$ at the acquired empirical distribution. As we now show, this is a simple calculation. Given a stationary $P_{\mathbf{Z}}$, $\delta \in (0, 1/2)$ and $l$, if $\delta$ is $l$-feasible there exists a corresponding stationary input process $\mathbf{X}$ which when passed through a $BSC(\delta)$ yields $P_{Z^l}$. Therefore, letting $\{C_j^{(l)}\}_{j=1}^{2^l}$ denote the set of binary

$l$-tuples, we can define and rewrite

$$\beta_j^{(l)} \triangleq P(Z_1 = 1 | Z_{-l+1}^0 = C_j^{(l)}) \tag{32}$$
$$= \frac{\sum_{r=1}^{2^l} (\delta * \alpha_r^{(l)}) P(x_{-l+1}^0 = C_r^{(l)}) \prod_{i=-l+1}^{0} \Pi_\delta(x_i, z_i = C_{j,i}^{(l)})}{P(Z_{-l+1}^0 = C_j^{(l)})}$$

where $*$ denotes binary convolution (defined as $p * q = (1-p)q + p(1-q)$), $\Pi_\delta$ is the channel matrix associated with the BSC($\delta$), $\Pi_\delta(x, z)$ is the channel transition probability, and

$$\alpha_j^{(l)} \triangleq P\left(X_1 = 1 | X_{-l+1}^0 = C_j^{(l)}\right) \quad \forall j.$$

For simplicity define,

$$\gamma_j^{(l)} \triangleq P\left(X_{-l+1}^0 = C_j^{(l)}\right)$$

and

$$\Theta_j^{(l)} \triangleq P\left(Z_{-l+1}^0 = C_j^{(l)}\right)$$

and let $\gamma$ and $\Theta$ be the associated length $2^l$ column vectors. Then,

$$\Theta^T = \gamma^T \Pi_\delta^{\otimes l}$$

where $\Pi_\delta^{\otimes l}$ denotes the $l^{th}$ tensor product of the matrix $\Pi_\delta$. Since $\delta < 1/2$, $\Pi_\delta^{-1}$ exists and we have

$$\gamma = (\Pi_\delta^{-T})^{\otimes l} \Theta.$$

Using the above notation, we can write

$$\beta_j^{(l)} = \frac{\sum_{r=1}^{2^l} (\alpha_r^{(l)} * \delta) [(\Pi_\delta^{-T})^{\otimes l} \Theta]_r \Psi_{r,j}}{\Theta_j},$$

where

$$\Psi = \Pi_\delta^{\otimes l}.$$

We can simplify the summation using vector notation. Dropping subscripts to indicate the corresponding vector gives

$$\beta_j^{(l)} \Theta_j = (\alpha^{(l)} * \delta)^T \left[(\Pi_\delta^{-T})^{\otimes l} \Theta\right] \odot \Psi_j,$$

where $\odot$ denotes component-wise multiplication and $\Psi_j$ is the $j^{th}$ column vector of $\Psi$. We can simplify further to obtain

$$\beta^{(l)} \odot \Theta = \Psi^T \left[((\Pi_\delta^{-T})^{\otimes l} \Theta) \odot (\alpha^{(l)} * \delta)\right]$$

which, following standard algebraic manipulations gives

$$\alpha^{(l)} * \delta = \left[(\Pi_\delta^{-T})^{\otimes l} (\beta^{(l)} \odot \Theta)\right] \oslash \left[(\Pi_\delta^{-T})^{\otimes l} \Theta\right], \tag{33}$$

where $\oslash$ denotes component-wise division. We will also use the following easily verified identity

$$\alpha^{(l)} = \frac{(\alpha^{(l)} * \delta) - \delta \bar{\mathbf{1}}}{1 - 2\delta}, \tag{34}$$

where $\bar{1}$ denotes the "all ones" column vector of appropriate dimension (in this case $2^l$). Combining (33) and (34) gives an explicit expression for $\alpha^{(l)}$ in terms of $\beta^{(l)}$.

With this notation in mind, $\Gamma_l$ is nothing but the maximum value of $\delta \in [0, 1/2]$ such that all the component of $\alpha^{(l)}$ are in $[0, 1]$. Hence a given $\delta$ is $l$-feasible if and only if all the components of the associated $\alpha^{(l)}$ are in $[0, 1]$. With each iteration of this $l$-feasibility test, we can shrink the uncertainty in $\Gamma_l$ by a factor of $1/2$. We can therefore quickly converge to the true $\Gamma_l$ as well as give precision bounds for a fixed number of iterations. In particular, neglecting empirical noise, after $T$ iterations of the $l$-feasibility test, we know our estimate is within $\pm 2^{-T-1}$ of $\Gamma_l$.

## V. ALGORITHM

Having discussed various aspects of computing the universal minimax denoiser in Sections III.C and IV.A, in this section we present the pseudo-code for the universal denoiser of Section III.A and discuss the complexity of the algorithm. As before, the size of the data is denoted by $n$ and the alphabet by $\mathcal{A}$. Algorithm: Minimax Denoiser

1  **initialize** Fix context length $k$, channel estimation parameter $l$, and channel estimation tolerance $T$
2  Construct empirical distributions, $P_{Z^k_{-k}}$ and $P_{Z^l}$
3  $\delta_U = \mathcal{U}$ and $\delta_L = 0$
4  **for** $i = 1$ to $T$
5    **if** $(\delta_U + \delta_L)/2$ is $l$-feasible
6      $\delta_L = (\delta_U + \delta_L)/2$
7    **elseif** $(\delta_U + \delta_L)/2$ is not $l$-feasible
8      $\delta_U = (\delta_U + \delta_L)/2$
9    **end**
10 **end**
11 $\hat{\Delta}_l = (\delta_U + \delta_L)/2$
12 $m_1 = \min_{f \in S_1(\hat{\Delta}_l)} B$
13 $f_1 = \arg\min_{f \in S_1(\hat{\Delta}_l)} B$
14 $m_2 = \min_{f \in S_2(\hat{\Delta}_l)} (A\hat{\Delta}_l^2 - A\hat{\Delta}_l - \hat{\Delta}_l + B)/(1 - 2\hat{\Delta}_l)$
15 $f_2 = \arg\min_{f \in S_2(\hat{\Delta}_l)} (A\hat{\Delta}_l^2 - A\hat{\Delta}_l - \hat{\Delta}_l + B)/(1 - 2\hat{\Delta}_l)$
16 $m_3 = \min_{f \in S(\hat{\Delta}_l)} (A\delta'^2 - A\delta' - \delta' + B)/(1 - 2\delta')$
17 $f_3 = \arg\min_{f \in S(\hat{\Delta}_l)} (A\delta'^2 - A\delta' - \delta' + B)/(1 - 2\delta')$
18 $\hat{X}^{n,k,l}(z^n) = f_i$ where $m_i = \min\{m_1, m_2, m_3\}$
19 Denoise using $\hat{X}^{n,k,l}(z^n)$

where $l$-feasible is defined in Section IV, $A$, $B$, and $\delta'$ are functions of $P_{Z^k_{-k}}$ and $f$ defined in Section III.C, and the minimizations are accomplished using convex optimization.

The complexity of the algorithm can be broken down as follows,

1) Collecting context data is $O(n)$ and requires $|\mathcal{A}|^{l+1}$ and $|\mathcal{A}|^{2k+1}$ memory cells

2) Constructing empirical distributions, $P_{Z_{-k}^k}$ and $P_{Z_l^0}$, requires $O(|\mathcal{A}|^{2k})$ and $O(|\mathcal{A}|^l)$ operations, respectively

3) Calculating $\hat{\Delta}_l$ estimate requires $O(|\mathcal{A}|^{2l})$ operations

4) Complexity of the convex optimization steps is roughly $O((|\mathcal{A}|^{2k+1})^3) = O(|\mathcal{A}|^{6k})$

5) Performing the actual denoising is $O(n)$

From Corollary 6 in [4], we note that $k$ and $l$ should not increase faster than $\log(n)/(16 \log |\mathcal{A}|)$. This implies that the algorithm requires memory of order $O(n)$ and that the complexity of the convex optimization and $l$-feasibility test is $O(n)$. In other words, the total complexity of algorithm is $O(n)$ and it requires $O(n)$ memory cells. From this we see that the algorithm scales gracefully in $n$. For a detailed discussion of the complexity of the convex optimization algorithm, see [1].

In comparison, the DUDE algorithm in [10] requires $O(n)$ computations, and memory of order $O(n)$. It should be noted that the DUDE does not require any convex optimization.

## VI. SIMULATIONS AND EXPERIMENTATION

In this section, we present experimental results obtained by implementation and employment of the scheme of Section III.A for the case of a binary signal corrupted by a BSC with an unknown crossover probability. We shall refer to the minimax scheme in this case as the Minimax Binary Denoiser (MBD), which we implement using the methods presented earlier. We compare the performance of the MBD to that of DUDE from [10] on simulated sequences (one dimensional), simulated fields (two dimensional), and on two real world images. For comparison, the DUDE of [10] is employed with the channel estimate suggested in [10], the one developed in Section IV, and the true channel parameter. Throughout the simulations, the channel crossover probability is, a priori, only known to lie in $[0, 1/2)$.

### A. A Modified DUDE

In [10, Section 8-C] the problem of denoising an unknown source, corrupted by an unknown discrete memoryless channel is considered. The algorithm suggested is to estimate the channel parameters and then to apply the DUDE assuming the channel estimate in lieu of the unknown channel parameters. The channel estimate suggested (as mentioned in Section IV) is,

$$\hat{\delta}_l = \min_{j,i} \min \left\{ \varphi_{j,i}^{(l)}, 1 - \varphi_{j,i}^{(l)} \right\}. \tag{35}$$

We refer to the application of the DUDE using a channel estimate as a Modified DUDE algorithm (M-DUDE). We propose an improvement to this algorithm: Rather than using the estimate in (35), which in general, as argued in Section IV, loosely upper bounds the largest feasible channel crossover probability, we suggest using the estimate $\hat{\Gamma}_l$ in (29), which, by (30), converges to the true upper bound.

### B. One Dimensional Simulations

We implemented the MBD as discussed in Section III for 1D sequences going through a BSC. As a source for our simulation we chose a hidden Markov source. To generate the source, a first-order symmetric binary Markov sequence with transition probability $p$ was simulated and then sent through a simulated BSC with parameter $\delta_s$. The hidden Markov source was then corrupted by a simulated BSC with parameter $\delta$, see Figure 1.

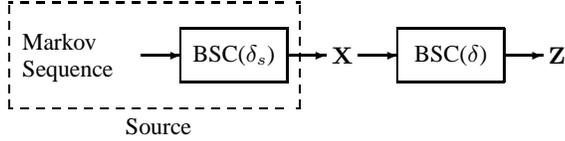

Fig. 1. The box diagram shows the process of creating the source, **X**, and the output, **Z** after going through a BSC with transition probability $\delta$.

We then compare the performance of the MBD to that of the M-DUDE. Since the MBD is a randomized denoiser, the actual denoising is performed by drawing a random variable, independently for each symbol, according to the distribution given by the MBD for the observed context. The output of the denoiser is then compared to the clean source, and the error is calculated. We apply the M-DUDE with two channel parameters: the estimate in (35), as suggested in [10], and our estimate from Section IV, $\hat{\Gamma}_l$. To get an idea of the optimum channel-dependent performance we also ran DUDE using the true value of the channel crossover probability $\delta$. Tables I and II show the results of the simulations.

| | | Estimate of $\delta$ | | Denoising Performance | | | |
|---|---|---|---|---|---|---|---|
| $\delta_s$ | $\delta$ | $\hat{\Gamma}_4$ | $\hat{\delta}_4$ | M-DUDE($\hat{\Gamma}_4$) | M-DUDE($\hat{\delta}_4$) | DUDE | MBD |
| 0 | 0.1000 | 0.1016 | 0.1320 | 0.0773 | 0.0773 | 0.0773 | 0.0775 |
| 0.0100 | 0.0918 | 0.1015 | 0.1328 | 0.0798 | 0.0798 | 0.0761 | 0.0776 |
| 0.0200 | 0.0833 | 0.0979 | 0.1319 | 0.0741 | 0.0820 | 0.0741 | 0.0741 |
| 0.0300 | 0.0745 | 0.1058 | 0.1336 | 0.0838 | 0.0838 | 0.0724 | 0.0771 |
| 0.0400 | 0.0652 | 0.1037 | 0.1333 | 0.0856 | 0.0856 | 0.0651 | 0.0762 |
| 0.0500 | 0.0556 | 0.1030 | 0.1333 | 0.0872 | 0.0872 | 0.0559 | 0.0754 |
| 0.0600 | 0.0455 | 0.1011 | 0.1335 | 0.0889 | 0.0889 | 0.0451 | 0.0745 |
| 0.0700 | 0.0349 | 0.0999 | 0.1314 | 0.0682 | 0.0902 | 0.0349 | 0.0682 |
| 0.0800 | 0.0238 | 0.1045 | 0.1320 | 0.0913 | 0.0913 | 0.0238 | 0.0711 |
| 0.0900 | 0.0122 | 0.1039 | 0.1317 | 0.0919 | 0.0919 | 0.0120 | 0.0690 |
| 0.1000 | 0 | 0.1078 | 0.1327 | 0.0927 | 0.0927 | 0 | 0.0674 |

TABLE I

DENOISING USING A HIDDEN MARKOV SOURCE WHERE THE MARKOV SOURCE HAS TRANSITION PROBABILITY $p = 0.15$. HERE $\delta_s * \delta$ IS FIXED AT 0.1, $k = 2$ AND SAMPLE SIZE IS $10^6$.

We note that when both the source is unknown, and there is channel uncertainty, there is a risk of injecting noise as the simulation results show. Where this is the case we notice that the MBD injects substantially fewer errors than the M-DUDE. This is a consequence of the worst case criterion the MBD was designed to optimize, which leads to more conservative denoising. We also note that in a number of cases the performance of the MBD is actually comparable to that of the channel-dependent DUDE (shown in [10] to achieve optimum source- and channel-dependent performance).

Finally, we note that the M-DUDE($\hat{\Gamma}_4$) performs consistently better than M-DUDE($\hat{\delta}_4$), often performing comparably to the channel-dependent DUDE. This is due to the fact that $\hat{\Gamma}$ is a better (asymptotically consistent) estimate

| | | Estimate of $\delta$ | | Denoising Performance | | | |
|---|---|---|---|---|---|---|---|
| $\delta_s$ | $\delta$ | $\hat{\Gamma}_4$ | $\hat{\delta}_4$ | M-DUDE($\hat{\Gamma}_4$) | M-DUDE($\hat{\delta}_4$) | DUDE | MBD |
| 0 | 0.0500 | 0.0560 | 0.0807 | 0.0430 | 0.0471 | 0.0430 | 0.0430 |
| 0.0050 | 0.0455 | 0.0567 | 0.0807 | 0.0425 | 0.0485 | 0.0425 | 0.0425 |
| 0.0100 | 0.0408 | 0.0618 | 0.0803 | 0.0505 | 0.0505 | 0.0412 | 0.0434 |
| 0.0150 | 0.0361 | 0.0617 | 0.0800 | 0.0513 | 0.0513 | 0.0359 | 0.0426 |
| 0.0200 | 0.0313 | 0.0620 | 0.0800 | 0.0527 | 0.0527 | 0.0313 | 0.0424 |
| 0.0250 | 0.0263 | 0.0580 | 0.0804 | 0.0404 | 0.0545 | 0.0262 | 0.0404 |
| 0.0300 | 0.0213 | 0.0587 | 0.0804 | 0.0398 | 0.0562 | 0.0213 | 0.0398 |
| 0.0350 | 0.0161 | 0.0518 | 0.0807 | 0.0390 | 0.0573 | 0.0160 | 0.0390 |
| 0.0400 | 0.0109 | 0.0569 | 0.0798 | 0.0380 | 0.0582 | 0.0109 | 0.0380 |
| 0.0450 | 0.0055 | 0.0560 | 0.0803 | 0.0372 | 0.0598 | 0.0054 | 0.0372 |
| 0.0500 | 0 | 0.0591 | 0.0810 | 0.0364 | 0.0612 | 0 | 0.0364 |

TABLE II

DENOISING USING A HIDDEN MARKOV SOURCE WHERE THE MARKOV SOURCE HAS TRANSITION PROBABILITY $p = 0.15$. HERE $\delta_s * \delta$ IS FIXED AT 0.05, $k = 2$ AND SAMPLE SIZE IS $10^6$.

of the largest feasible $\delta$, while $\hat{\delta}$ (even asymptotically) is in most cases a strict upper bound to it.

*C. Two Dimensional Data*

We begin with simulations similar to those of the previous section, except now on a two-dimensionally-indexed simulated process (image) as the source. We implemented the MBD for 2D sources as in [6], and compared it to a 2D implementation of the DUDE for a BSC as in [11]. As before, we ran the M-DUDE with the estimates suggested in [10] found in (35), and our estimate from section IV, $\hat{\Gamma}_l$. As in the previous section, for the source- and channel-dependent optimum performance benchmark we also ran DUDE using the true $\delta$.

Here the context used for denoising consists of the $3 \times 3$ square centered on the symbol to be denoised. Furthermore, for calculating $\hat{\Gamma}_3$ and $\hat{\delta}_3$, we use the 3 bits in the upper left-hand corner of the $3 \times 3$ square.

For our experimentation we use a hidden random field (HRF). First we generate a binary random field causally by letting each pixel component depend stochastically on the pixel to its left and the pixel above it, where the left and top boundary are drawn according to a Bernoulli distribution with parameter $1/2$. More specifically,

$$x_{i,j} = \begin{cases} N_{i,j}^{1/2} & \text{if} \quad i = 0 \text{ or } j = 0 \\ N_{i,j}^{1/2} & \text{if} \quad x_{i-1,j} \neq x_{i,j-1} \\ N_{i,j}^{\alpha} & \text{if} \quad x_{i-1,j} = x_{i,j-1} = 0 \\ N_{i,j}^{\bar{\alpha}} & \text{if} \quad x_{i-1,j} = x_{i,j-1} = 1, \end{cases} \quad (36)$$

where $x_{i,j}$ denotes the component at location $(i,j)$ and $\{N_{i,j}^{1/2}\}$, $\{N_{i,j}^{\alpha}\}$ and $\{N_{i,j}^{\bar{\alpha}}\}$ are independent fields, consisting of independent components which are Bernoulli with parameters $1/2$, $\alpha$, and $\bar{\alpha} = 1 - \alpha$, respectively. We then corrupt this field by a BSC with transition probability $\delta_s$ and the output is then used as the noiseless image for the simulation, i.e., analogously as in Figure 1, replacing the Markov sequence by a random field.

The hidden random field is then sent through a BSC with transition probability $\delta$ and denoising is performed. We used a test image of size $2000 \times 2000$. Tables III and IV show the bit error rate of the denoised image relative to the noiseless one.

The results show a trend similar to that observed for one-dimensional signals. We notice that the MBD consistently outperforms the M-DUDE($\hat{\delta}_3$) and is comparable in performance to the DUDE with the true channel parameter

|       |      | Estimate of $\delta$ | | Denoising Performance | | | |
|-------|------|---------|---------|---------------------|----------------------|--------|--------|
| $\delta_s$ | $\delta$ | $\hat{\Gamma}_3$ | $\hat{\delta}_3$ | M-DUDE($\hat{\Gamma}_3$) | M-DUDE($\hat{\delta}_3$) | DUDE | MBD |
| 0 | 0.01 | 0.0184 | 0.0606 | 0.0115 | 0.0182 | 0.0095 | 0.0114 |
| 0 | 0.02 | 0.0279 | 0.0717 | 0.0169 | 0.0225 | 0.0169 | 0.0169 |
| 0 | 0.05 | 0.0575 | 0.105  | 0.0354 | 0.0457 | 0.0354 | 0.0354 |
| 0 | 0.10 | 0.106  | 0.160  | 0.0682 | 0.0696 | 0.0687 | 0.0691 |

TABLE III

DENOISING RESULTS FOR A HIDDEN RANDOM FIELD WITH $\alpha = 0.05$, IMAGE SIZE $2000 \times 2000$.

|       |      | Estimate of $\delta$ | | Denoising Performance | | | |
|-------|------|---------|---------|---------------------|----------------------|--------|--------|
| $\delta_s$ | $\delta$ | $\hat{\Gamma}_3$ | $\hat{\delta}_3$ | M-DUDE($\hat{\Gamma}_3$) | M-DUDE($\hat{\delta}_3$) | DUDE | MBD |
| 0.0500 | 0.0556 | 0.1019 | 0.1187 | 0.0619 | 0.0677 | 0.0556 | 0.0550 |
| 0.0750 | 0.0294 | 0.101  | 0.119  | 0.0757 | 0.0836 | 0.0295 | 0.0542 |
| 0.0250 | 0.0263 | 0.0504 | 0.0639 | 0.0335 | 0.0337 | 0.0263 | 0.0628 |
| 0.0375 | 0.0135 | 0.0499 | 0.0640 | 0.0416 | 0.0428 | 0.0137 | 0.0268 |
| 0.0150 | 0.0052 | 0.0207 | 0.0313 | 0.0156 | 0.0163 | 0.0051 | 0.0112 |
| 0.010  | 0.0102 | 0.0202 | 0.0314 | 0.0122 | 0.0135 | 0.0102 | 0.0122 |

TABLE IV

DENOISING RESULTS FOR HIDDEN RANDOM FIELD WITH $\alpha = 0.01$, IMAGE SIZE $2000 \times 2000$.

$\delta$, (shown to be optimal in [10]). Again, the M-DUDE($\hat{\Gamma}_3$) performs better than M-DUDE($\hat{\delta}_3$), and in fact does essentially as well as the MBD. As before, this is due to the fact that when the underlying data contains strong structure such as in the case of random fields, then $\hat{\Gamma}_l$ tends to be close to the true channel parameter $\delta$. This fact suggests that the M-DUDE($\hat{\Gamma}_l$) would be a good algorithm for denoising natural images, as is indeed observed in the examples that follow.

We next present denoising results for a binary text image. We scanned half a page of text at a resolution of $1000 \times 600$. In Table V we show a piece (approx. $1/6$) of the original, noisy, and denoised images for $\delta = 0.1$. For this particular case the DUDE using the true channel parameter $\delta$ had a normalized error rate of $0.0330$, while the MBD had an error rate of $0.0479$. The M-DUDE($\hat{\delta}_3$) did better than the MBD, with error rate close to that of the DUDE with the true channel parameter. This should be contrasted with Tables III and IV, where the MBD consistently did better than the M-DUDE($\hat{\delta}_3$). The high performance of the M-DUDE($\hat{\delta}_3$) is explained by the fact that we were denoising a text image. A text image is composed of mostly white background which allows for any reasonable channel estimate to be highly accurate. Hence in this case the M-DUDE($\hat{\delta}_3$) is effectively implementing the DUDE with the true channel parameter, the optimal denoiser. On the other hand, the MBD takes a more conservative approach believing that the true channel crossover probability could be anywhere between $0$ and $\hat{\Gamma}_3$.

Our final result is the denoising of a binary image. We used a halftoned image of size $1200 \times 1785$. In Table VI we show the original, noisy, and denoised images at approximately $1/6$ scale. In this case, the channel parameter $\delta$ is set to $0.02$. As before, we implemented the DUDE using three channel parameters, the true channel parameter $\delta$, $\hat{\delta}_3$, and $\hat{\Gamma}_3$. We also denoised using the MBD. The normalized errors of the denoisers were, $0.019$ for the DUDE with the true channel parameter, $0.0327$ for the M-DUDE($\hat{\delta}_3$), $0.0289$ for the M-DUDE($\hat{\Gamma}_3$), and $0.019$ for the MBD. Notice that both the M-DUDE($\hat{\delta}_3$) and M-DUDE($\hat{\Gamma}_3$) inject errors. This is not the case for the MBD which reduces the amount of errors. Hence, unlike the text images of Table V, the MBD outperforms the DUDE for

| TABLE V |
|---|

DENOISING OF A TEXT IMAGE. THE TOP IMAGE IS THE ORIGINAL IMAGE, NEXT IS THE NOISY VERSION WHERE $\delta = 0.1$, THEN A DENOISED VERSION USING THE DUDE WITH THE TRUE $\delta$ AND FINALLY A DENOISED VERSION USING MBD.

both channel estimates. The reason for this difference is due to the fact that with images incorporating fine detail, it is hard if not impossible to estimate the channel parameter. Stated otherwise, the original image is inherently noisy. Therefore denoisers that are optimized for a particular channel, such as the DUDE, may perform poorly and, in some cases, inject errors. Even with this difficulty to estimate the channel parameter, we observe that the M-DUDE($\hat{\Gamma}_3$) outperformed the M-DUDE($\hat{\hat{\delta}}_3$). This is because, as shown in previous results, $\hat{\Gamma}_l$ tends to be closer to the true maximally consistent channel parameter than $\hat{\hat{\delta}}_l$. In particular, for this case, $\hat{\Gamma}_3$ was $0.0524$ and $\hat{\hat{\delta}}_3$ was $0.0683$. Also observe that, similarly to the hidden random field results of Tables III and IV, the MBD does as well as the DUDE with the true channel parameter.

## VII. Beyond Binary Alphabets

Up to this point we have confined attention to binary alphabets. Our goal in this section is to extend the algorithms developed in Sections III.C and IV.A to the more general non-binary setting. In particular, we will address the case of non-binary alphabets with channels parameterized by a single parameter. In other words, the uncertainty in the knowledge of the channel can be described by the uncertainty of a single parameter, analogously as in the binary case where the uncertainty was in the the crossover probability of the $BSC$ (knowing that the crossover probability is less than $1/2$). We also require some structure in this free parameter, $\delta$. We require that the channels be monotonic in $\delta$, i.e., that if $\delta_0$ is feasible for a given $P_\mathbf{Z}$ then all $0 \leq \delta \leq \delta_0$ are also feasible. As before, we say that $\delta$ is feasible if there exists a source such that, when passed through the channel defined by $\delta$, the output statistics agree with $P_\mathbf{Z}$.

Observe that with the above constraint, our definition of $\Gamma(P_\mathbf{Z})$, extended from the binary case in (5), becomes

$$\Gamma(P_\mathbf{Z}) = \max \{\delta : \exists P_\mathbf{X} \text{ s.t. } P_\mathbf{X} * Ch(\delta) = P_\mathbf{Z}\}, \tag{37}$$

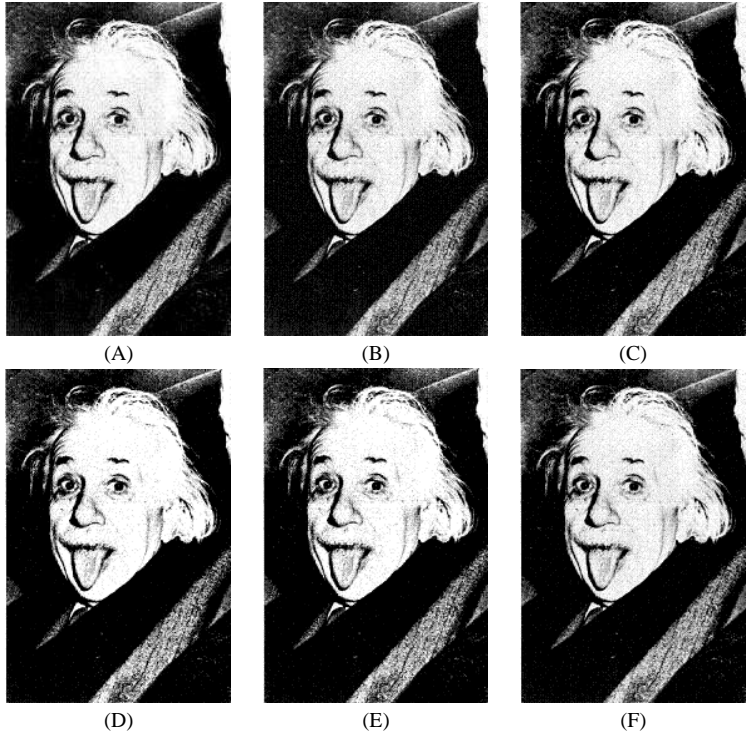

TABLE VI

DENOISING OF A HALFTONE BINARY IMAGE. IMAGE (A) IS THE ORIGINAL IMAGE, (B) IS THE NOISY VERSION WHERE $\delta = 0.02$, (C) IS THE IMAGE DENOISED USING THE DUDE WITH TRUE CHANNEL PARAMETER, WITH NORMALIZED ERROR OF $= 0.019$, (D) IS THE IMAGE DENOISED USING THE DUDE WITH CHANNEL ESTIMATE $\hat{\hat{\delta}}_3$, WITH NORMALIZED ERROR OF $0.0327$, (E) IS THE IMAGE DENOISED USING THE DUDE WITH CHANNEL ESTIMATE $\hat{\Gamma}_3$, WITH NORMALIZED ERROR OF $0.0289$, AND (F) IS THE IMAGE DENOISED USING THE MBD, WITH NORMALIZED ERROR OF $0.019$.

where $P_{\mathbf{X}} * Ch(\delta)$ denotes the distribution of the output process of the channel with the parameter $\delta$ whose input process has distribution $P_{\mathbf{X}}$. We define for this setting a minimax performance benchmark, similarly to (9),

$$\mu(P_{\mathbf{Z}}, \mathbf{Z}) = \lim_{k \to \infty} \limsup_{n \to \infty} \min_{f \in \mathcal{F}_k} \mathcal{L}_f^{(n)}(P_{\mathbf{Z}}, \mathbf{Z}), \tag{38}$$

where here $\mathcal{L}_f^{(n)}$ is the loss of the sliding window scheme $f$, extended from the binary setting.

By confining our channel uncertainty into a single parameter, we can preserve much of the structure from the binary alphabet case. With this structure intact we can adapt the algorithms from Sections III.C and IV.A to the single parameterized non-binary alphabet case. This adaptation needs to be done on a case by case basis. In the following section we illustrate how this is done for a particular family of non-binary channels. We denote the alphabet by $\mathcal{A} = \{1, \ldots, M\}$, i.e. $M$ is the alphabet size. The family of channels we will use for our example can be considered the M-ary generalization of the BSC.

## A. The Symmetric Channel

Let the probability transition matrix of the channel be given by

$$\begin{pmatrix} 1-\delta & \frac{\delta}{M-1} & \cdots & \frac{\delta}{M-1} \\ \frac{\delta}{M-1} & \ddots & \ddots & \vdots \\ \vdots & \ddots & \ddots & \frac{\delta}{M-1} \\ \frac{\delta}{M-1} & \cdots & \frac{\delta}{M-1} & 1-\delta \end{pmatrix}.$$

It is readily verified that if a single variable $X \in \mathcal{A}$ is distributed as

$$P_X = \begin{pmatrix} \frac{(M-1)\alpha_1 - \delta}{(M-1) - M\delta} \\ \vdots \\ \frac{(M-1)\alpha_{M-1} - \delta}{(M-1) - M\delta} \\ \frac{(M-1) - \delta - (M-1)\sum_{i=1}^{M-1} \alpha_i}{(M-1) - M\delta} \end{pmatrix}$$

is corrupted by the above channel then the channel output distribution is

$$P_Z = (\alpha_1, \ldots, \alpha_{M-1}, 1 - \sum_{i=1}^{M-1} \alpha_i)^T. \tag{39}$$

We represent a denoiser for this single-observation problem as $\{d_{i,j}\}_{i,j=1}^M$, with $d_{i,j}$ denoting the probability that the denoiser outputs a reconstruction $j$ upon observing $i$. In particular $d_{i,j} \in [0,1]$ and $\sum_{j=1}^M d_{i,j} = 1\ \forall i$. The expected loss of this denoiser (in the single observation problem) is

$$\begin{aligned} F\left(\{\alpha_i\}_{i=1}^M, \delta, \{d_{i,j}\}_{i,j=1}^M\right) = \\ \frac{(M-1)\alpha_1 - \delta}{(M-1) - M\delta} \left[(1-\delta)\sum_{j \neq 0} d_{0,j} + \frac{\delta}{M-1} \sum_{i \neq 0, j \neq 0} d_{i,j}\right] + \\ \frac{(M-1)\alpha_2 - \delta}{(M-1) - M\delta} \left[(1-\delta)\sum_{j \neq 1} d_{1,j} + \frac{\delta}{M-1} \sum_{i \neq 1, j \neq 1} d_{i,j}\right] + \ldots + \\ \frac{(M-1) - \delta - (M-1)\sum_{i=1}^{M-1}\alpha_i}{(M-1) - M\delta} \left[(1-\delta)\sum_{j \neq M} d_{M,j} + \frac{\delta}{M-1} \sum_{i \neq M, j \neq M} d_{i,j}\right] = \\ \sum_{k=1}^{M-1} \frac{(M-1)\alpha_k - \delta}{(M-1) - M\delta} \left[(1-\delta)\sum_{j \neq k} d_{k,j} + \frac{\delta}{M-1} \sum_{i \neq k, j \neq k} d_{i,j}\right] + \\ \frac{(M-1) - \delta - (M-1)\sum_{i=1}^{M-1}\alpha_i}{(M-1) - M\delta} \left[(1-\delta)\sum_{j \neq M} d_{M,j} + \frac{\delta}{M-1} \sum_{i \neq M, j \neq M} d_{i,j}\right], \end{aligned} \tag{40}$$

where $\alpha_i$ is defined in (39).

Consider now a probability distribution on $M$-ary $(2k+1)$-tuples, $P_{Z_{-k}^k}$. Consider a denoiser for one symbol based on observing a noise-corrupted $(2k+1)$-tuple around it $f : \mathcal{A}^{2k+1} \to \mathcal{S}(\mathcal{A})$. Such a denoiser can be thought of as a collection of single-symbol denoisers, one for each context, where if the current context is $z_{-k}^{-1}, z_1^k$ and we observe $i$ as the middle symbol, the denoiser will change the symbol to $j$ with probability $f([z_{-k}^{-1}, i, z_1^k])[j]$. We define the following functional

$$G_k\left(P_{Z_{-k}^k}, \delta, f\right) = \sum_{z_{-k}^{-1}, z_1^k \in \{1,2,...,M\}^k} F\left(\left\{P_{Z_0|z_{-k}^{-1}, z_1^k}(Z_0 = i)\right\}_{i=1}^M, \delta, \{f([z_{-k}^{-1}, i, z_1^k])\}_{i=1}^M\right) P_{Z_{-k}^k}(z_{-k}^{-1}, z_1^k), \tag{41}$$

where $P_{Z_0|z_{-k}^{-1},z_1^k}$ denotes $\Pr(Z_0 = 1|Z_{-k}^{-1} = z_{-k}^{-1}, Z_1^k = z_1^k)$ under the source $P_{Z_{-k}^k}$. We define

$$f_{MM_k}\left[P_{Z_{-k}^k}, \Delta\right] = \arg\min_f \max_{0 \leq \delta \leq \Delta} G_k\left(P_{Z_{-k}^k}, \delta, f\right), \tag{42}$$

selecting an arbitrary achiever when it is not unique. Let $\hat{X}^{n,k,l}$ denote the $n$-block denoiser defined by,

$$\hat{X}_{[i]}^{n,k,l}(z^n) = f_{\mathbb{MM}_k}\left(\hat{Q}^{2k+1}[z^n], \hat{\Delta}_l(z^n)\right)[z_{i-k}^{i+k}] \quad k+1 \leq i \leq n-k, \tag{43}$$

### B. Algorithms for the Symmetric Channel

Now that we have examined the general symmetric channel, we want to extend the algorithms from sections III.C and IV.A to this more general case. We observe first that the single parameter structure discussed earlier exists in the case of the symmetric channel. Combined with the invertibility of the channel matrix, this single parameter structure is all we need to adapt the algorithm in section IV.A to the generalized symmetric channel case. The same algebraic construction found in section IV.A follows with the modification that

$$\beta_{j,i}^{(l)} \triangleq P(Z_1 = i|Z_{-l+1}^0 = C_j^{(l)}) \tag{44}$$

$$\alpha_{j,i}^{(l)} \triangleq P(X_1 = i|X_{-l+1}^0 = C_j^{(l)}). \tag{45}$$

Hence the channel described by a particular $\delta$ is $l$-feasible if and only if each row of the associated matrix $\alpha^{(l)}$ is an element of the M-dimensional simplex. Therefore we have an algorithm that not only converges to $\Delta_l$, but also produces bounds with each iteration.

Now that we have extended the algorithm from section IV.A, we turn our attention to solving (42). As in (12) we define $J_k$ by

$$J_k\left(P_{Z_{-k}^k}, \Delta, f\right) = \max_{0 \leq \delta \leq \Delta} G_k\left(P_{Z_{-k}^k}, \delta, f\right). \tag{46}$$

Similarly to the analysis in Section III.C, from (40) and (41) it follows that

$$G_k\left(P_{Z_{-k}^k}, \delta, f\right) = \frac{A\delta^2 + B\delta + C}{M - 1 - M\delta}$$

where $A(P_{Z_{-k}^k}, f)$, $B(P_{Z_{-k}^k}, f)$, and $C(P_{Z_{-k}^k}, f)$ are affine functions and can be derived as in the binary case of Section III.C. Therefore $J_k$ can be expressed as

$$\max\left\{C, \frac{A\Delta^2 + B\Delta + C}{M - 1 - M\Delta}, \frac{A\delta'^2 + B\delta' + C}{M - 1 - M\delta'}\mathbf{1}_{\delta' \in (0,\Delta)}, \frac{A\delta''^2 + B\delta'' + C}{M - 1 - M\delta''}\mathbf{1}_{\delta'' \in (0,\Delta)}\right\}, \tag{47}$$

where

$$\delta' = \frac{AM - A + \sqrt{A(AM^2 - 2AM + A + BM^2 - BM - CM^2)}}{AM} \quad \text{and}$$

$$\delta'' = \frac{AM - A - \sqrt{A(AM^2 - 2AM + A + BM^2 - BM - CM^2)}}{AM}.$$

Hence, as in Section III.C, $J_k$ simplifies to the max between four points which are simple functions of the coefficients $A$, $B$ and $C$. Hence, for a given denoiser $f$, the quantity $J_k(f)$ is easily calculated.

The analysis for the minimization of $J_k$ for the simple binary case in section III.C is quite involved. Certain subtleties in the analysis suggest that the general $M$-array symmetric channel may not be piecewise convex or that finding the boundaries of the convex regions may be overly complicated. Hence this analysis needs to be carried out and verified for the particular alphabet size at hand.

Since we cannot simply extend the algorithm from Section III.C to the general $M$-array symmetric channel case without analysis for each $M$ of interest, is there something that can be done in general? Earlier it was shown that for a given denoiser $f$ we can easily calculate $J_k(f)$. This suggests that although it might not be possible to find the absolute minimum of $J_k$, we can apply methods such as simulated annealing to estimate the absolute minimum, see [8] and [9] for a detailed discussion of simulated annealing over a continuous domain. Both the simulated annealing methods discussed in [8] and [9] are concerned with unconstrained minimization. Since we are dealing with the constrained minimization of $J_k$, we once again will need to make use of the log barrier method as described in [1]. The benefit of using simulated annealing is that one can estimate the minimax optimal denoiser and have some control over the complexity versus accuracy of the estimation. The control over the trade off comes from controlling the annealing schedule.

We have therefore managed to extend the algorithm from Section IV.A to the general $M$-ary symmetric channel case. We have also shown that for a particular $M$, it is possible to extend the algorithm from Section III.C to the general $M$-ary symmetric channel case, and that even if one cannot use the convex optimization methods developed in Section III.C, estimates of the minimax optimal denoiser can still be obtained using simulated annealing. Hence, in practice, for any $M$, one can apply the minimax denoiser for the $M$-ary symmetric channel.

## C. Beyond Symmetric Channels

The analysis and methods used in Sections VII.A and VII.B can be extended to many other families of channels. With use of simulated annealing, see [8] and [9], the above methods can be applied to any family of channels with the proper scalar parameterization and with easily calculated expressions for $J_k$. This significantly extends the possible applications of the minimax denoiser to the non-binary case.

## VIII. MINIMAX $\neq$ MAXIMIN

A natural question arising in the context of our minimax criterion is whether it coincides with the maximin. An affirmative answer would imply that a minimax optimal scheme is an optimal scheme for the least denoisable source-channel pair consistent with the output distribution. This, in turn, would suggest that employing the DUDE of [10] tailored for the least denoisable source-channel pair, which can easily be estimated, would give rise to a universally minimax optimal and practical scheme.

Unfortunately, as we now show, the minimax does not coincide with the maximin in our problem. Specifically, we shall argue that for some noise-corrupted sources the minimax is greater than the maximin. To show this we assume the input process, $P_\mathbf{X}$ is Bernoulli $p$, $p < 1/2$, and is corrupted by a $BSC(\delta)$ channel. We assume that the channel crossover probability, $\delta$, belongs to an uncertainty set $\mathcal{U}$, which is given to us. In particular, assume that $\mathcal{U} = \{0, 0.01, 0.02, \ldots, 0.49\}$. As a performance measure we take the unconditional expected loss

$$\min_{f \in \mathcal{F}_k} \max_{\{(P_\mathbf{X}, \delta) : \delta \in \mathcal{U}, P_\mathbf{X} * \delta = P_\mathbf{Z}\}} E_{[P_\mathbf{X}, \delta]}[L_f(X^n, Z^n)]. \quad (48)$$

Corollary 2 of [4] ensures that the minimax denoiser, as defined in Section III.A, attains the minimum in (48) and hence justifies the use of the unconditional measure.

With this setup, the output process, $P_\mathbf{Z}$, will be Bernoulli with parameter $p * \delta$, where $\delta \in \mathcal{U}$, and we can assume it is known to us. In such a case, there are only two optimal schemes, as well as mixtures of the two. Depending on the channel parameter $\delta$, the optimal scheme either outputs what it sees ($\delta < p$) or outputs all zeros ($\delta > p$). The transitional point is when $\delta = p$.

We now consider the $(\delta, p)$ pair such that $\delta = p$ and $p * \delta = \Pr\{Z = 1\}$; Denote the value of $\delta$ and $p$ that satisfy this by $\delta^*$. For this case it can be shown that the scheme attaining the minimum in (48) assigns a unique probability $a \in (0, 1)$ to the "say what you see" scheme and $\bar{a}$ to the "say all zeros" scheme. In other words it is a mixture of the two optimal schemes. In order for the minimax to be equivalent to the maximin, the optimal minimax denoiser, in the sense of (48), would have to be an optimal denoiser for the worst possible channel, i.e. to swap the minimax for maximin, the channel maximizing the loss would also need to be a $BSC$ with transition probability $\delta^*$. Figure 2 is a plot of $\delta^*$ and the $\delta$ maximizing the loss with respect to $\Pr\{Z = 1\}$, denoted by $\delta_w$. In other words, $\delta_w$ is the least denoisable channel which agrees with $P_\mathbf{Z}$.

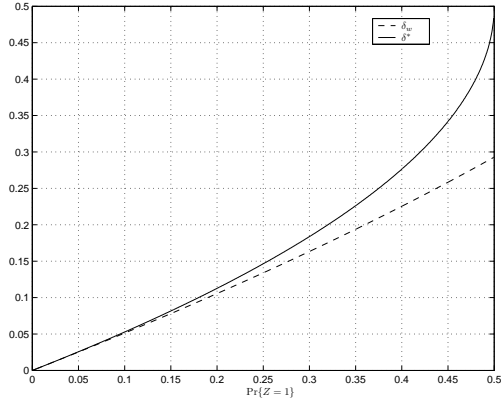

Fig. 2. Plot of $\delta^*$ and $\delta_w$ with respect to the parameter of the Bernoulli noise-corrupted source.

Evidently, it is not the case that $\delta^* = \delta_w$. Hence, with Figure 2 in mind, we can easily verify that for $P_\mathbf{Z}$ equal to Bernoulli(0.18), $\delta^* = 0.1$,

$$\min_{f \in M_{opt}} \max_{\{(P_\mathbf{X}, \delta) : \delta \in \mathcal{U}, P_\mathbf{X} * \delta = P_\mathbf{Z}\}} E_{[P_\mathbf{X}, \delta]}[L_f(X^n, Z^n)] > \max_{\{(P_\mathbf{X}, \delta) : \delta \in \mathcal{U}, P_\mathbf{X} * \delta = P_\mathbf{Z}\}} \min_{f \in M_{opt}} E_{[P_\mathbf{X}, \delta]}[L_f(X^n, Z^n)] \quad (49)$$

where $M_{opt}$ is the set of denoisers which are optimal for some $(P_\mathbf{Z}, \delta)$ pair, i.e., a denoiser is an element of $M_{opt}$ if there exists some pair $(P_\mathbf{Z}, \delta)$ for which it is optimal. This observation leads us to the following:

*Theorem 2 (Minimax $\neq$ Maximin):* There exist stationary ergodic sources $P_{\mathbf{Z}}$ for which

$$\min_{f \in \mathcal{F}_k} \max_{\{(P_{\mathbf{X}}, \delta): \delta \in \mathcal{U}, P_{\mathbf{X}} * \delta = P_{\mathbf{Z}}\}} E_{[P_{\mathbf{X}}, \delta]}[L_f(X^n, Z^n)] > \max_{\{(P_{\mathbf{X}}, \delta): \delta \in \mathcal{U}, P_{\mathbf{X}} * \delta = P_{\mathbf{Z}}\}} \min_{f \in \mathcal{F}_k} E_{[P_{\mathbf{X}}, \delta]}[L_f(X^n, Z^n)] \qquad (50)$$

**Proof:** Assume, by contradiction, that (50) does not hold when $P_{\mathbf{Z}}$ is the Bernoulli$(0.18)$ source. This implies that the minimizing denoiser is optimal for the worst possible channel. Hence to attain the minimax performance it is enough to minimize only over mixtures of *optimal* denoisers, i.e.,

$$\min_{f \in \mathcal{F}_k} \max_{\{(P_{\mathbf{X}}, \delta): \delta \in \mathcal{U}, P_{\mathbf{X}} * \delta = P_{\mathbf{Z}}\}} E_{[P_{\mathbf{X}}, \delta]}[L_f(X^n, Z^n)] = \min_{f \in M_{opt}} \max_{\{(P_{\mathbf{X}}, \delta): \delta \in \mathcal{U}, P_{\mathbf{X}} * \delta = P_{\mathbf{Z}}\}} E_{[P_{\mathbf{X}}, \delta]}[L_f(X^n, Z^n)]. \qquad (51)$$

Combining (51) with (49) gives

$$\min_{f \in \mathcal{F}_k} \max_{\{(P_{\mathbf{X}}, \delta): \delta \in \mathcal{U}, P_{\mathbf{X}} * \delta = P_{\mathbf{Z}}\}} E_{[P_{\mathbf{X}}, \delta]}[L_f(X^n, Z^n)] > \max_{\{(P_{\mathbf{X}}, \delta): \delta \in \mathcal{U}, P_{\mathbf{X}} * \delta = P_{\mathbf{Z}}\}} \min_{f \in M_{opt}} E_{[P_{\mathbf{X}}, \delta]}[L_f(X^n, Z^n)]. \qquad (52)$$

On the other hand clearly

$$\max_{\{(P_{\mathbf{X}}, \delta): \delta \in \mathcal{U}, P_{\mathbf{X}} * \delta = P_{\mathbf{Z}}\}} \min_{f \in M_{opt}} E_{[P_{\mathbf{X}}, \delta]}[L_f(X^n, Z^n)] \geq \max_{\{(P_{\mathbf{X}}, \delta): \delta \in \mathcal{U}, P_{\mathbf{X}} * \delta = P_{\mathbf{Z}}\}} \min_{f \in M_k} E_{[P_{\mathbf{X}}, \delta]}[L_f(X^n, Z^n)] \qquad (53)$$

which when combined with (52) implies that (50) holds, contradicting our assumption.

□

As shown in Corollary 2 of [4], the minimax denoiser has the property of attaining the minimum in (48). However, Theorem 2 implies that such a denoiser is not in general an optimal denoiser for the worst source-channel pair. Therefore, the minimax denoiser and an optimal denoiser for the worst source-channel pair are not in general the same. Accordingly, a denoiser designed to be optimal for the worse source-channel pair is not guaranteed to be minimax optimal.

## IX. CONCLUSIONS

In [4], denoisers that are asymptotically optimal in a worst case sense are suggested for the setting of an unknown source corrupted by a DMC, under channel uncertainty. The present paper was dedicated to the implementation of these denoisers. We have presented efficient algorithms for implementing the denoisers suggested in [4] for the binary alphabet as well as for efficiently estimating the set of feasible channels in the uncertainty set. We also extended these algorithms to a large family of channels in the non-binary case, focusing on the generalized M-ary symmetric channel. It was shown that the suggested universally min-max denoisers do not correspond to schemes that attain the optimum distribution-dependent performance under the worst case source-channel pair. In general, the min-max denoisers are not optimal distribution-dependent schemes for any source-channel pair, implying that use of the DUDE of [10] with a channel estimate is suboptimal under the worst case loss criterion. We have also presented a natural modification to the original DUDE, M-DUDE($\hat{\Gamma}_l$), which employs the DUDE using an estimate of channel parameter, which was shown to be consistent in the sense of converging to the largest feasible channel parameter. Simulations shown suggest that, in practice, this scheme may perform well in denoising images under channel uncertainty, often attaining performance which is comparable to that of the MBD.

ACKNOWLEDGMENTS

The authors are grateful to Marcelo Weinberger for constructive comments that helped to improve the manuscript.